\documentstyle[twocolumn,psfig]{article}
\newtheorem{defi}{Definition}[section]

\newtheorem{lemma}{Lemma}[section]

\newtheorem{theo}{Theorem}[section]
\newtheorem{prot}{Protocol}[section]

\makeatletter
\def\@normalsize{\@setsize\normalsize{12pt}\xpt\@xpt
\abovedisplayskip 10pt plus2pt minus5pt\belowdisplayskip \abovedisplayskip
\abovedisplayshortskip \z@ plus3pt\belowdisplayshortskip 6pt plus3pt
minus3pt\let\@listi\@listI}

\def\subsize{\@setsize\subsize{12pt}\xipt\@xipt}

\def\section{\@startsection {section}{1}{\z@}{24pt plus 2pt minus 2pt}
{12pt plus 2pt minus 2pt}{\large\bf}}

\def\subsection{\@startsection {subsection}{2}{\z@}{12pt plus 2pt minus 2pt}
{12pt plus 2pt minus 2pt}{\subsize\bf}}
\makeatother

\begin{document}


\date{}

\title{\Large\bf Quantum $m$-out-of-$n$ Oblivious Transfer\thanks{
This work is partially supported by a grant from the Ministry of
Science and Technology (\#2001CCA03000), National Natural Science
Fund (\#60273045) and Shanghai Science and Technology Development
Fund (\#03JC14014).}}


\author{ Zhide Chen,\
      \ Hong Zhu \\[.2cm]
\small{}\textit{Department of Computer Science, Fudan University, Shanghai 200433, P.R.China.} \\[.1cm]
\small{}\textit{Key Laboratory of Intelligent Information
Processing, Fudan University, Shanghai 200433, P.R.China.} \\[.1cm]
 \small{} \{02021091,  hzhu\}@fudan.edu.cn }

\maketitle

\thispagestyle{empty}

\subsection*{\centering Abstract}
{\em In the $m$-out-of-$n$ Oblivious Transfer ($OT$) model, one
party $Alice$ sends $n$ bits to another party $Bob$, $Bob$ can get
only $m$ bits from the $n$ bits. However, $Alice$ cannot know
which $m$ bits $Bob$ received. Y.Mu and Naor presented classical
$m$-out-of-$n$ Oblivious Transfer based on discrete logarithm. As
the work of Shor, the discrete logarithm can be solved in
polynomial time by quantum computers, so such $OT$s are unsecure
to the quantum computer. In this paper, we construct a quantum
$m$-out-of-$n$ $OT$ ($QOT$) scheme based on the transmission of
polarized light and show that the scheme is robust to general
attacks, i.e. the $QOT$ scheme satisfies statistical correctness
and statistical privacy.\\
\\
\textbf{Keywords.} \ Quantum, Oblivious Transfer. }

\section{Introduction}
A number of recent papers have provided compelling evidence that
certain computational, cryptographic, and information theoretic
tasks can be performed more efficiently by models based on quantum
physics than those based on classical physics~\cite{[Shor97]}.
\par Oblivious Transfer (OT) is used as a key component in
many applications of cryptography~\cite{[WIE],[EGL85],[R81]}.
Informally speaking in an Oblivious Transfer, $Alice$ sends a bit
to $Bob$ that he receives half the time (this fact is out of their
control), $Alice$ does not find out what happened, $Bob$ knows if
he get the bit or nothing. Similarly, in a 1-out-of-2 Oblivious
Transfer, $Alice$ has two bits $b_0,b_1$ that she sends to  $Bob$
in such a way that he can decide to get either of them at his
choosing but not both. $Alice$ never finds out which bit $Bob$
received.
\par In 2001, Naor presented a 1-out-of-n Oblivious Transfer~\cite{[Naor01]},
Y.Mu showed that $m$-out-of-$n$ Oblivious Transfer could also be
realized based on the discrete logarithm. In the $m$-out-of-$n$
Oblivious Transfer($1\leq m<n$) , $Alice$ sends $n$ bits to $Bob$,
$Bob$ can get only $m$ of them. In the case of quantum, Claude
Cr\'{e}peau provided a 1-out-of-2 quantum Oblivious Transfer based
on the transmission of polarized light in 1994. The protocol of
Cr\'{e}peau's can be used directly to implement a one-out-of-three
Oblivious Transfer.
\par The organization of this paper is as following: in section 2,
we give the definitions of the correctness and privacy of the
$m$-out-of-$n$ OT protocol. In section 3, we review the 1-out-of-2
OT of Claude Cr\'{e}peau and its intuition. In section 4, we
construct an $m$-out-of-$n$ OT, and in section 5 we show that this
scheme satisfies statistical correctness and statistical privacy .

\section{Definitions}
The natural constraints(see below) of correctness and privacy of a
$m$-out-of-$n$ OT($1\leq m<n$) is showed below.
\begin{defi} \textbf{Perfect Correctness:} It should be that when $Alice$
and $Bob$ follow the protocol and start with $Alice's$ input bits
$b_{1},b_{2},\cdots,b_{n}$ and $Bob's$ input $c_1,c_2,\dots,c_m\in
\{1,2,\cdots,n\}$, they finish with $Bob$ getting
$b_{c_1},b_{c_2},\cdots,b_{c_m} \in $ $\{$ $b_{1}$, $b_{2}$,
$\cdots$, $b_{n}\}$.
\end{defi}

\begin{defi}
\textbf{Perfect Privacy:} It should be that, $Alice$ can not find
out about $c_1,c_2,\dots,c_m$, and $Bob$ can not find out more
than $m$ of $b_1,b_2,\dots,b_n$.
\end{defi}
\par The protocol we describe in the next section is of
probabilistic nature. We cannot show that this protocol perfectly
satisfies the above constraints but satisfies in a statistical
sense: after an amount of work in $O(N)$ time the protocol will
satisfy for some positive constant $\epsilon <1$.

\begin{defi}\textbf{Statistical
Correctness:} It should be that , except with probability at most
$\varepsilon^{N}$, when $Alice$ and $Bob$ follow the protocol and
start with $Alice's$ input bits $b_1,b_2,\dots,b_n$ and $Bob's$
input $c_1,c_2,\dots,c_m\in \{1,2,\cdots,n\}$ they finish with
$Bob$ getting $b_{c_1},b_{c_2},\cdots,b_{c_m}\in
\{b_{1},b_{2},\cdots,b_{n}\}$.
\end{defi}

\begin{defi}
\textbf{Statistical Privacy:} It should be that, except with
probability at most $\epsilon^{N}$, $Alice$ can not find out
$c_1,c_2,\dots,c_m$, and $Bob$ can not find out  more than $m$ of
$b_1,b_2,\dots,b_n$.
\end{defi}

\section{Quantum 1-out-of-2 Oblivious Transfer}
In this section, we introduce the quantum 1-out-of-2 OT provided
by Claude Cr\'{e}peau~\cite{[C94]}. Let $\copyright\!\!\!\!|$ \ \
denote the random variable that takes the binary value 0 with
probability 1/2 and 1 with probability 1/2. Also, denote by
$[\quad]_{i}$ the selection function such that
$[a_{0},a_{1},\cdots,a_{k}]_{i}=a_{i}$. Let $\leftrightarrow
\!\!\!\!\updownarrow\
=(|\!\!\!\leftrightarrow\rangle,|\!\!\uparrow\!\!\!\downarrow\rangle)$
and
$\nwarrow\!\!\!\!\!\!\searrow\!\!\!\!\!\!\nearrow\!\!\!\!\!\!\swarrow\
=(|\!\!\nwarrow\!\!\!\!\!\!\searrow\rangle,|\!\nearrow\!\!\!\!\!\!\swarrow
\rangle)$ denote respectively the bases of rectilinear and
diagonal polarization in the quantum state space of a photon. The
quantum 1-out-of-2 OT is as follows:

\subsection{Quantum 1-out-of-2 OT}
\begin{prot} 1-out-of-2 OT$(b_{0},b_{1})(c)$
\begin{enumerate}
    \item $DO_{i=1}^{2n}$
        \begin{itemize}
            \item $Alice$ picks a random bit $r_{i} \leftarrow \copyright\!\!\!\!|$
            \item $Alice$ picks a random bit $\beta_{i}\leftarrow
            \copyright\!\!\!\!|$ and defines her emission
            basis
            $(|\varphi_{i}\rangle, |\varphi_{i}^{\perp}\rangle)\leftarrow
            [\leftrightarrow \!\!\!\!\updownarrow,\nwarrow\!\!\!\!\!\!\searrow\!\!\!\!\!\!\nearrow\!\!\!\!\!\!\swarrow]_{\beta_{i}} $
            \item $Alice$ sends to Bob a photon $\pi_{i}$ with polarization $[|\varphi_{i}\rangle, |\varphi_{i}^{\perp}\rangle]_{r_i}$
            \item $Bob$ picks a random bit  $\beta'_i \leftarrow
            \copyright\!\!\!\!|$ and measures $\pi_{i}$ in basis $(|\theta_{i}\rangle, |\theta_{i}^{\perp}\rangle)\leftarrow[\leftrightarrow \!\!\!\!\updownarrow,\nwarrow\!\!\!\!\!\!\searrow\!\!\!\!\!\!\nearrow\!\!\!\!\!\!\swarrow]_{\beta_{i}'} $
            \item $Bob$ sets $r_{i}' \leftarrow
                \left\{%
                \begin{array}{ll}
                    0, & \hbox{if $\pi_{i}$ is observed as $|\theta_{i}\rangle$} \\
                    1, & \hbox{if $\pi_{i}$ is observed as $|\theta_{i}^{\bot}\rangle$} \\
                \end{array}%
                \right.$
        \end{itemize}

    \item $DO_{i=1}^{n}$
    \begin{itemize}
            \item Bob runs
            $commit(r'_{i})$, $commit(\beta'_{i})$, $commit(r'_{n+i})$, $commit(\beta'_{n+i})$
            with $Alice$
            \item $Alice$ picks $c_{i} \leftarrow \copyright\!\!\!\!|$ \ and
            announces it to $Bob$
            \item Bob runs $unveil(r'_{nc_{i}+i}), unveil(\beta'_{nc_{i}+i})$
            \item $Alice $ checks that $\beta_{nc_{i}+i}=\beta'_{nc_{i}+i}\rightarrow r_{nc_{i}+i}=r'_{nc_{i}+i}$
            \item if $c_{i}=0$ then $Alice$ sets
            $\beta_{i}\leftarrow\beta_{n+i}$ and $r_{i}\leftarrow r_{n+i}$  and $Bob$ set
            $\beta'_{i}\leftarrow\beta'_{n+i}$ and $r'_{i}\leftarrow r'_{n+i}$
    \end{itemize}

    \item $Alice$ announces her choices
    $\beta_{1}\beta_{2}\cdots\beta_{n}$ to $Bob$
    \item $Bob$ randomly selects two subsets $I_{0},I_{1}\subset
            \{1,2,\cdots,n\}$ subject to $|I_{0}|=|I_{1}|=n/3$, $I_{0} \cap
            I_{1}=\emptyset$ and $\forall i \in I_{c},
            \beta_{i}=\beta_{i}'$, and he announces $\langle I_{0},I_{1}
            \rangle$ to $Alice$
    \item $Alice$ receives $\langle J_{0},J_{1}  \rangle$=$\langle I_{0},I_{1}
            \rangle$, computes and sends $\widehat{b}_{0}\leftarrow b_{0} \oplus \bigoplus_{j \in J_{0}}r_{j}$ and  $\widehat{b}_{1}\leftarrow b_{1} \oplus \bigoplus_{j \in J_{1}}r_{j} $
    \item $Bob$ receives $\langle \widehat b_{0},\widehat b_{1}
    \rangle$ and computes $b_{c}\leftarrow \widehat {b}_{c} \oplus \bigoplus_{j \in J_{c}} r'_{j} $
\end{enumerate}

\end{prot}

\subsection{Intuition behind 1-out-of-2 OT}
\par In this 1-out-of-2 QOT, $Alice$ must prevent $Bob$ from storing
the photons and waiting until she discloses the bases before
measuring them, which would allow him to obtain both of $Alice's$
bits with certainty. To realize this, $Alice$ gets $Bob$ to
$commit$ to the bits that he received and the bases that he used
to measure them. Before going ahead with $r_i$, say, $Alice$
checks that $Bob$ had committed properly to $r_{n+i}$ when he read
that bit in the basis that she used to encode it. If at any stage
$Alice$ observes a mistake ($\beta_{n+i}=\beta'_{n+i}$ but
$r_{n+i}\neq r'_{n+i}$), she stops further interaction with $Bob$
who is definitely not performing his legal protocol (this should
never happen if $Bob$ follows his protocol).
\par
In this protocol, $r_{1}r_{2} \cdots r_{n}$ are chosen by $Alice$
in step 1 and are sent to $Bob$ via an ambiguous coding referred
to as the BB84 coding~\cite{[BB84]}: when $Alice$ and $Bob$ choose
the same emission and reception basis, the bit received is the
same as what was sent and uncorrelated otherwise. $Bob$ builds two
subsets: one $I_{c}$ that will allow him to get $b_{c}$, and one
$I_{\overline{c}}$ that will spoil $b_{\overline{c}}$. The
calculations of steps 5-6 are much that all the bits in a subset
must be known by $Bob$ in order for him to be able to obtain the
output bit connected to that subset.

\section{Protocol for Quantum $m$-out-of-$n$ Oblivious Transfer}
\subsection{Weak Bit Commitment}
In 1993, Gilles Brassard, etc provided a quantum bit commitment
scheme provably unbreakable by both parties~\cite{[BCJL]}.
However, unconditionally quantum bit commitment was showed
impossible~\cite{Mayers}. In~\cite{[DAUA]}, Aharonov provided a
weak bit commitment.
\begin{defi}~\cite{[DAUA]}
In the weak bit commitment protocol, the following requirements
should hold.
\begin{itemize}
\item If both Alice and Bob are honest, then  both Alice and Bob
accept.

\item (Binding) If Alice tries to change her mind about the value
of $b$, then there is non zero probability that an honest Bob
would reject.

\item (Sealing) If Bob attempts to learn information about the
deposited bit $b$, then there is non zero probability that an
honest Alice would reject.
\end{itemize}
\end{defi}
In the following scheme, $Bob$ will use this weak quantum bit
commitment to commit.
\subsection{Intuition for $m$-out-of-$n$ OT}
\par In the $m$-out-of-$n$ OT, $Bob$ should build $n$ subsets
$I_{1},I_{2},\dots,I_{n}\subseteq \{1,2,\cdots,n\}$, $m$ of that
will allow him to get $b_{c_1},b_{c_2},\dots,b_{c_m}$
($c_1,c_2,\dots,c_m \in \{ 1,2,\dots,n \}$), and the other $I$'s
will spoil the remnant $b$'s. In $I_1\cup I_2\cup\cdots\cup I_n$,
the rate of the $i$'s satisfying $\beta_{i}'=\beta_{i}$ would be
more than $\frac{m}{n}$ and less than $\frac{m+1}{n}$. i.e.
$$\frac{m}{n}\leq \frac{\# \{i| \beta _{i}= \beta _{i}', i\in I_1\cup\cdots\cup I_n\}}{|I_1\cup\cdots\cup I_n|}<\frac{m+1}{n} $$
In our scheme, we let the rate to be
$\frac{\frac{m}{n}+\frac{m+1}{n}}{2}=\frac{2m+1}{2n}$. As
$\beta$'s and $\beta'$'s are choice randomly, we have
$$\lim_{N\rightarrow \infty} \frac{\#\{ \beta_i=\beta_i' \}}{N}=\frac{1}{2}.$$
For a large $N$, the rate of $i$'s in $\{1,2,\cdots,N \}$ that
satisfy $\beta_{i}'=\beta_{i}$ would be approximately
$\frac{1}{2}$, then $Bob$ should remove some $i$'s from the
$\{1,2,\cdots,N \}$. The number of $i$'s that should be removed
can be calculated as following:
\\If $\frac{2m+1}{2n}<\frac{1}{2}$, there are more $i$'s that satisfy $\beta_{i}'=\beta_{i}$ than
required, so $Bob$ should remove $x$ $i$'s  that satisfying
$\beta_{i}'= \beta_{i}$ from $\{1,2,\cdots,N \}$. $x$ can be
calculated as follows:
\begin{eqnarray*}
   \frac{\frac{N}{2}-x}{N-x}&=&\frac{2m+1}{2n}\\
   x&=&\frac{n-(2m+1)}{2n-(2m+1)}N
\end{eqnarray*}
 If $\frac{2m+1}{2n}\geq\frac{1}{2}$, there are more $i$'s
that satisfy $\beta_{i}'\neq\beta_{i}$ than required, so $Bob$
should remove $x$ $i$'s  that satisfying $\beta_{i}'\neq
\beta_{i}$ from $\{1,2,\cdots,N \}$. $x$ can be calculated as
follows:
\begin{eqnarray*}
\frac{\frac{N}{2}}{N-x}&=&\frac{2m+1}{2n}\\
x&=&\frac{(2m+1)-n}{2m+1}N
\end{eqnarray*}
$N$ must satisfy $(2n-(2m+1))(2m+1)|((2m+1)-n)N$ so that $x$ would
be an interger. we let the $i$'s that was removed from
$\{1,2,\cdots,N\}$ be $u_1,u_2,\cdots,u_x$.

\subsection{Quantum $m$-out-of-$n$ OT}
In the $m$-out-of-$n$ $QOT$, $Alice$ has input
$b_1,b_2,\cdots,b_n$, $Bob$ has input $c_1,c_2,\cdots,c_m$. The
output of the scheme is $b_{c_1},b_{c_2},\cdots,b_{c_m}$.

\begin{prot} $m$-out-of-$n$ QOT$(b_1,b_2,\dots,b_n)(c_1,c_2,\dots,c_m)$
\begin{enumerate}
    \item $DO_{i=1}^{2N}$
        \begin{itemize}
            \item $Alice$ picks a random bit $r_{i} \leftarrow \copyright\!\!\!\!|$
            \item $Alice$ picks a random bit $\beta_{i}\leftarrow
            \copyright\!\!\!\!|$\ \  and defines her emission basis
            $(|\varphi_{i}\rangle, |\varphi_{i}^{\perp}\rangle)\leftarrow[\leftrightarrow \!\!\!\!\updownarrow,\nwarrow\!\!\!\!\!\!\searrow\!\!\!\!\!\!\nearrow\!\!\!\!\!\!\swarrow]_{\beta_{i}} $
            \item $Alice$ sends to Bob a photon $\pi_{i}$ with polarization $[|\varphi_{i}\rangle, |\varphi_{i}^{\perp}\rangle]_{r_i}$
            \item $Bob$ picks a random bit  $\beta'_i \leftarrow
            \copyright\!\!\!\!|$\ \  and measures $\pi_{i}$ in basis
             $(|\theta_{i}\rangle, |\theta_{i}^{\perp}\rangle)\leftarrow[\leftrightarrow \!\!\!\!\updownarrow,\nwarrow\!\!\!\!\!\!\searrow\!\!\!\!\!\!\nearrow\!\!\!\!\!\!\swarrow]_{\beta_{i}'} $
            \item $Bob$ sets $r_{i}' \leftarrow
                \left\{%
                \begin{array}{ll}
                    0, & \hbox{if $\pi_{i}$ is observed as $|\theta_{i}\rangle$} \\
                    1, & \hbox{if $\pi_{i}$ is observed as $|\theta_{i}^{\bot}\rangle$} \\
                \end{array}%
                \right.    $
        \end{itemize}

    \item $DO_{i=1}^{N}$
    \begin{itemize}
            \item Bob runs
            $commit(r'_{i})$, $commit(\beta'_{i})$, $commit(r'_{N+i})$, $commit(\beta'_{N+i})$
            with $Alice$
            \item $Alice$ picks $d_{i} \leftarrow \copyright\!\!\!\!|$ \ \ and
            announces it to $Bob$
            \item Bob runs $unveil(r'_{Nd_{i}+i}), unveil(\beta'_{Nd_{i}+i})$
            \item $Alice $ checks that $\beta_{Nd_{i}+i}=\beta'_{Nd_{i}+i}\rightarrow r_{Nd_{i}+i}=r'_{Nd_{i}+i}$
            \item if $d_{i}=0$ then $Alice$ sets
            $\beta_{i}\leftarrow\beta_{N+i}$ and $r_{i}\leftarrow r_{N+i}$  and $Bob$ set
            $\beta'_{i}\leftarrow\beta'_{N+i}$ and $r'_{i}\leftarrow r'_{N+i}$
    \end{itemize}

    \item $Alice$ announces her choices
    $\beta_{1}\beta_{2}\cdots\beta_{N}$ to $Bob$

    \item $DO_{j=1}^{x}$
    \begin{itemize}
            \item If $\frac{2m+1}{2n}<\frac{1}{2}$ Bob runs
            $unveil(r'_{u_j})$, $
            unveil(\beta'_{u_j})$ that satisfying
                    $\beta_{u_j}= \beta'_{u_j}$, $Alice $ checks that $\beta_{u_j}=\beta'_{u_j}\rightarrow r_{u_j}=r'_{u_j}$
            \item If $\frac{2m+1}{2n}\geq \frac{1}{2}$ Bob runs
            $unveil(r'_{u_j})$, $unveil(\beta'_{u_j})$ that satisfying
                    $\beta_{u_j}\neq \beta'_{u_j}$

    \end{itemize}
    \item $Bob$ randomly selects n subsets $I_{1},I_{2},\cdots,I_{n}   \subset
            \{1,2,\cdots,N\}-\{ u_1,u_2,\dots,u_x \}$ subject to
            $|I_{1}|=|I_{2}|=\cdots=|I_{n}|=(N-x)/n$, $\forall j\neq k$, $ I_{j} \cap
            I_{k}=\emptyset$ and $\forall j$ $\in I_{c_1}\cup I_{c_2}\cup \cdots\cup
            I_{c_m}$, $\beta_{j}=\beta_{j}'$, and he announces $\langle
            I_{1},I_{2},\cdots,I_{n}
            \rangle$ to $Alice$
    \item $Alice$ receives $\langle J_{1},J_{2},\cdots ,J_{n} \rangle$=$\langle
            I_{1},I_{2},\cdots,I_{n}  \rangle$,    computes and sends $\widehat{b}_{1}\leftarrow b_{1} \oplus \bigoplus_{j \in J_{1}}r_{j}$, $\widehat{b}_{2}\leftarrow b_{2} \oplus \bigoplus_{j \in
    J_{2}}r_{j}$, $\cdots$, $\widehat{b}_{n}\leftarrow b_{n} \oplus \bigoplus_{j \in
    J_{n}}r_{j}$ to $Bob$
    \item $Bob$ receives $\langle \widehat b_{1},\widehat
    b_{2},\cdots, \widehat b_{n}
    \rangle$ and computes $b_{c_i}\leftarrow \widehat {b}_{c_i} \oplus \bigoplus_{j \in J_{c_i}}
    r'_{c_j} , $ $i=1,2,\cdots,m$
\end{enumerate}

\end{prot}

\section{Analysis}
In the $m$-out-of-$n$ $QOT$, $Bob$ must read the photons sent by
$Alice$ as they come: he cannot wait and read them later,
individually or together. We assume that the channel used for the
quantum transmission is free of errors, so that it is guaranteed
that $r_{i}'=r_{i}$ whenever $\beta_{i}'=\beta_{i}$. we now show
that under the assumption this protocol satisfies the statistical
version of the above constraints.

\subsection{Correctness}

\begin{lemma} $\mathbf{Hoefding \quad inequality}$~\cite{[HO]}
Let $X_1, X_2,\cdots, X_n$ be total independent random variables
with identical probability distribution so that $E(X_i)=\mu$ and
the range of $X_i$ is in $[a,b]$. Let the simple average
$Y=(X_1+X_2+\cdots+X_n)/n$ and $\delta>0$, then
$$Pr[|Y-\mu|\geq \delta]\leq
2\cdot e^{\frac{-2n\cdot \delta^2}{b-a}}$$

\end{lemma}

So, if $Pr[X_i=0]=Pr[X_i=1]=\frac{1}{2}$, then $\mu=\frac{1}{2}$
and $a=0,b=1$, we have the following inequality
$$Pr[|\sum _{i=1}^{n}\frac{X_i}{n}- \frac{1}{2}  |\geq \delta]\leq
2\cdot e^{-2\cdot n \delta^2}$$
\par We show that most of the time the output is correct if the
parties abide to their prescribed protocol. In a given run of the
protocol, $Bob$ will succeed in computing
$b_{c_1},b_{c_2},\dots,b_{c_m}$ properly provided satisfying the
following conditions :
\\ when $\frac{2m+1}{2n}< \frac{1}{2}$
$$\# \{i| \beta _{i}= \beta
_{i}'\}-x \geq  (N-x)m/n$$ or when $\frac{2m+1}{2n}\geq
\frac{1}{2}$
$$ \# \{i| \beta _{i}= \beta _{i}'\} \geq (N-x)m/n$$
Because in that case he can form $I_{c_1},I_{c_2},\dots,I_{c_m}$
as prescribed and then he can compute the output bit as
$\widehat{b}_{c_i}\oplus \bigoplus _{j\in I_{c_i} }r_{j}'$ which
is $$\widehat{b}_{c_i}\oplus \bigoplus _{j\in I_{c_i} }r_{j}'
=b_{c_i}\oplus \bigoplus _{j\in J_{c_i} }r_{j} \bigoplus _{j\in
I_{c_i} }r_{j}'=b_{c_i}\oplus \bigoplus _{j\in I_{c_i} }r_{j}
\oplus r_{j}'$$ because $J_{c_i}$ is $I_{c_i}$. Since
$\beta_{i}=\beta_{i}'\rightarrow r_{j} \oplus r_{j}'=0 $ makes all
the right terms vanish, we end up with
$$  \widehat{b}_{c_i}\oplus \bigoplus _{j\in I_{c_i}
}r_{j}' =b_{c_i} $$ Therefore the protocol gives the correct
output unless satisfying the following conditions : \\when
$\frac{2m+1}{2n}< \frac{1}{2}$
$$ \# \{i| \beta _{i}= \beta
_{i}'\}-x <  (N-x)m/n
$$ or when $\frac{2m+1}{2n}\geq \frac{1}{2}$
$$ \# \{i| \beta _{i}= \beta _{i}'\} <  (N-x)m/n  $$
in which case $Bob$ is unable to form the set
$I_{c_1},I_{c_2},\dots,I_{c_m}$ as prescribed. Now, we can
calculate the probability that $Bob$ can not form
$I_{c_1},I_{c_2},\dots,I_{c_m}$
\\If $\frac{2m+1}{2n}< \frac{1}{2}$ (i.e. $2m+1< n$,
$x=\frac{n-(2m+1)}{2n-(2m+1)}N$), then the probability that $Bob$
can get less than $m$ bits is given by
\begin{eqnarray*}
& & P[\# \{i| \beta _{i}= \beta _{i}'\}-x < (N-x)m/n]\\
&=& P[\# \{i| \beta _{i}= \beta _{i}'\} < (N-x)m/n+x]\\
 &=& P[\sum_{i=1}^{N} \beta _{i} \oplus \beta _{i}'
 >  N-((N-\frac{n-(2m+1)}{2n-(2m+1)}N)m/n\\
&&
 +\frac{n-(2m+1)}{2n-(2m+1)}N)]\\
&=& P[\frac{1}{N}\sum_{i=1}^{N} \beta _{i} \oplus \beta _{i}'
 >  1-\frac{n-(m+1)}{2n-(2m+1)}]\\
&=& P[\frac{1}{N}\sum_{i=1}^{N} \beta _{i} \oplus \beta _{i}'
 >  \frac{n-m}{2n-(2m+1)}]\\
&\leq & P[ | \frac{1}{N}\sum_{i=1}^{N} \beta _{i} \oplus \beta
_{i}'-\frac{1}{2}|> \frac{n-m}{2n-(2m+1)}-\frac{1}{2} ]
\end{eqnarray*}
It is easy to check that $\frac{n-m}{2n-(2m+1)}-\frac{1}{2}>0$.\\
Given that $P[\beta _{i} \oplus \beta _{i}'=1]=1/2$, let
$N>\frac{\ln 2}{(\frac{n-m}{2n-(2m+1)}-\frac{1}{2})^2}$, this
probability can be easily bounded by
\begin{eqnarray*}
&<& 2\cdot e^{-2\cdot N(\frac{n-m}{2n-(2m+1)}-\frac{1}{2})^2}\\
&=& 2\cdot e^{- N(\frac{n-m}{2n-(2m+1)}-\frac{1}{2}) ^2}
\cdot e^{- N(\frac{n-m}{2n-(2m+1)}-\frac{1}{2}) ^2} \\
&<& e^{- N(\frac{n-m}{2n-(2m+1)}-\frac{1}{2}) ^2}\\
&=&\varepsilon^N
\end{eqnarray*}
($\varepsilon= e^{-(\frac{n-m}{2n-(2m+1)}-\frac{1}{2})^2}<1$)
using Hoefding's inequality.
\\If $\frac{2m+1}{2n} \geq \frac{1}{2}$ (i.e. $2m+1\geq n$,
$x=\frac{(2m+1)-n}{2m+1}$), then the probability that $Bob$ can
get less than $m$ bits is given by
\begin{eqnarray*}
& & P[\# \{i| \beta _{i}= \beta _{i}'\} < (N-x)m/n]\\
&=& P[\sum_{i=1}^{N} \beta _{i} \oplus \beta _{i}'
 >  N-(N-\frac{(2m+1)-n}{2m+1}N)m/n]\\
&=& P[\frac{1}{N}\sum_{i=1}^{N} \beta _{i} \oplus \beta _{i}'
 >  1-\frac{m}{2m+1}]\\
&\leq & P[ | \frac{1}{N}\sum_{i=1}^{N} \beta _{i} \oplus \beta
_{i}'-\frac{1}{2}|>\frac{1}{2}- \frac{m}{2m+1} ]
\end{eqnarray*}
It is easy to check that $\frac{1}{2}- \frac{m}{2m+1}>0$.\\
Given
that $P[\beta _{i} \oplus \beta _{i}'=1]=1/2$, let $N>\frac{\ln
2}{(\frac{1}{2}-\frac{m}{2m+1}) ^2}$, this probability can be
easily bounded by
\begin{eqnarray*}
&<& 2\cdot e^{-2\cdot N  ( \frac{1}{2}-\frac{m}{2m+1}) ^2  }\\
&=& 2\cdot e^{- N  ( \frac{1}{2}-\frac{m}{2m+1}) ^2  }\cdot e^{- N  ( \frac{1}{2}-\frac{m}{2m+1}) ^2  }\\
&<& e^{- N  ( \frac{1}{2}-\frac{m}{2m+1}) ^2  } \\
&=&\varepsilon^N
\end{eqnarray*}
 ($\varepsilon=e^{-(\frac{1}{2}-\frac{m}{2m+1})^2}<1$) using Hoefding's
inequality.
\\So, $Bob$ can get less than $m$ bits that sent from $Alice$
with probability less than $\varepsilon^N$.
\subsection{Privacy}
We analyse the privacy of each party individually as if he or she
is facing a malicious opponent.

\subsubsection{Privacy for $Bob$}
\begin{theo}
$Alice$ can not find out much about $c_1,c_2,\dots,c_m$,
\end{theo}
\smallskip
\upshape \noindent \textit{Proof.}  The only things $Alice$ gets
though the protocol are the sets $J_1,J_2,\dots,J_n$.
$\beta_{i}$'s and $\beta_{i}'$'s are independent from each other.
$J_1,J_2,\dots,J_n$ will have uniform distribution over all
possible pairs of disjoint subsets of size $\frac{N-x}{n}$ for
$i=1,i=2,\dots$ as well as for $i=n$. Therefore $Alice$ learns
nothing about the $c_1,c_2,\dots,c_m$. $\hfill \Box$

\subsubsection{Privacy for $Alice$}
\begin{theo}
Except with probability at most $\epsilon^{n}$, $Bob$ can not find
out much information about more that $m$ of $b_1,b_2,\dots,b_n$.
\end{theo}

\smallskip
\upshape \noindent \textit{Proof.}  The probability of that $Bob$
gets more than $m$ bits (i.e. get at least $m+1$ bits). So
\\
\\If $\frac{2m+1}{2n}<\frac{1}{2}$ (i.e. $2m+1<n$,
$x=\frac{n-(2m+1)}{2n-(2m+1)}N$), the probability that $Bob$ can
get more than $m+1$ bits is given by
\begin{eqnarray*}
&& P[\# \{i| \beta _{i}= \beta _{i}'\} - x\geq (N-x)(m+1)/n]\\
&=& P[\# \{i| \beta _{i}= \beta _{i}'\} \geq (N-x)(m+1)/n+x]\\
 &=& P[\sum_{i=1}^{N} \beta _{i} \oplus \beta _{i}'
 \leq  N-((N-\frac{n-(2m+1)}{2n-(2m+1)}N)(m\\
&& +1)/n+\frac{n-(2m+1)}{2n-(2m+1)}N)]\\
&=& P[\frac{1}{N}\sum_{i=1}^{N} \beta _{i} \oplus \beta _{i}'
 \leq  1-\frac{n-m}{2n-(2m+1)}]\\
&\leq & P[ | \frac{1}{N}\sum_{i=1}^{N} \beta _{i} \oplus \beta
_{i}'-\frac{1}{2}|>\frac{1}{2}- \frac{n-m}{2n-(2m+1)} ]
\end{eqnarray*}
It is easy to check that $\frac{1}{2}- \frac{n-m}{2n-(2m+1)}>0$.\\
 Given that $P[\beta _{i} \oplus \beta _{i}'=1]=1/2$, let
 $N>\frac{\ln 2}{(\frac{1}{2}- \frac{n-m}{2n-(2m+1)}) ^2}$, this
probability can be easily bounded by
\begin{eqnarray*}
&<& 2\cdot e^{-2\cdot N ( \frac{1}{2}- \frac{n-m}{2n-(2m+1)}) ^2 }\\
&=& 2\cdot e^{-N ( \frac{1}{2}- \frac{n-m}{2n-(2m+1)}) ^2 }\cdot e^{- N ( \frac{1}{2}- \frac{n-m}{2n-(2m+1)}) ^2 }\\
&<& e^{- N ( \frac{1}{2}- \frac{n-m}{2n-(2m+1)}) ^2}\\
&=&\varepsilon^N
\end{eqnarray*}
($\varepsilon=e^{-(\frac{1}{2}- \frac{n-m}{2n-(2m+1)})^2}<1$)
using Hoefding's inequality.
\\
\\If $\frac{2m+1}{2n}\geq \frac{1}{2}$ (i.e. $2m+1\geq n$,
$x=\frac{(2m+1)-n}{2m+1}$), then the probability that $Bob$ can
get more than $m+1$ bits is given by
\begin{eqnarray*}
&& P[\# \{i| \beta _{i}= \beta _{i}'\} \geq (N-x)(m+1)/n]\\
&=& P[\sum_{i=1}^{N} \beta _{i} \oplus \beta _{i}'
 \leq  N-(N-\frac{(2m+1)-n}{2m+1}N)(m\\
&&+1)/n]\\
&=& P[\frac{1}{N}\sum_{i=1}^{N} \beta _{i} \oplus \beta _{i}'
 \leq  1-\frac{m+1}{2m+1}]\\
&\leq & P[ | \frac{1}{N}\sum_{i=1}^{N} \beta _{i} \oplus \beta
_{i}'-\frac{1}{2}|> \frac{m+1}{2m+1}-\frac{1}{2} ]
\end{eqnarray*}
It is easy to check that $\frac{m+1}{2m+1}-\frac{1}{2}>0$.\\
 Given that $P[\beta _{i} \oplus \beta _{i}'=1]=1/2$, let
 $N>\frac{\ln 2}{(\frac{m+1}{2m+1}-\frac{1}{2}) ^2}$,  the
probability can be easily bounded by
\begin{eqnarray*}
&<& 2\cdot e^{-2\cdot N ( \frac{m+1}{2m+1}-\frac{1}{2}) ^2  }\\
&=& 2\cdot e^{-N ( \frac{m+1}{2m+1}-\frac{1}{2}) ^2  }\cdot e^{- N ( \frac{m+1}{2m+1}-\frac{1}{2}) ^2  }\\
&<& e^{- N ( \frac{m+1}{2m+1}-\frac{1}{2}) ^2  }\\
&=&\varepsilon^N
\end{eqnarray*}
($\varepsilon=e^{-(\frac{m+1}{2m+1}-\frac{1}{2})^2}<1$) using
Hoefding's inequality.

\par Finally, we show that $Bob$ cannot get more than $m$ bits by
attacking the weak quantum bit commitment. Let the probability
that he can cheat $Alice$ in the weak QBC be $p$ ($0<p<1$), the
probability that he can get one more bit is
$p^{\frac{N-x}{n}}<\epsilon^N$ ($\epsilon=p^{\frac{1}{2n}}$).
\par So, $Bob$ can get more than $m$ bits that sent from $Alice$
with probability less than $\varepsilon^N$.

$\hfill \Box$

In the 1-out-of-2 OT scheme, $n=2$ and $m=1$,
$\frac{2m+1}{2n}=\frac{3}{4}>\frac{1}{2}$, then the probability is
less than
$$2\cdot e^{-N \cdot 2(
\frac{m+1}{2m+1}-\frac{1}{2}) ^2  }=2\cdot e^{-N \cdot 2(
\frac{2}{3}-\frac{1}{2}) ^2  }=2\cdot e^{-\frac{N}{18} }$$

\section{Conclusions and Future Work}
In this paper, we construct an quantum $m$-out-of-$n$ OT based on
the transmission of polarized light, which is an extension of the
quantum 1-out-f-2 OT, and prove that this scheme satisfies
statistical correctness and statistical privacy, i.e. except with
a small probability $\epsilon^N$, $Bob$ can get the correct $m$
bits, and cannot get one more bit than required.
\par
We think the following points is interesting for further research:

\begin{enumerate}
    \item Implement and apply the QOT in the real world.
    \item Find a QOT satisfies perfect correctness and perfect
    privacy.
\end{enumerate}


\begin{thebibliography}{99}
\bibitem{[BB84]}    Bennett, C.H. and Brassard, G., ``Quantum Cryptography: Public-key Distribution and Coin
Tossing``, In Proceedings of the International Conference on
Computers, Systems and Signal Processing, Bangalore, India,
December 1984, pp. 175-179.
\bibitem{[BCJL]}  Brassard, G., Cr\'{e}peau, C. Jozsa, R. and
Langlois, D., ``A Quantum Bit Commitment Scheme Probably
unbreakable by both parties``, In Proceedings of the 34th Annual
IEEE Symposium on Foundations of Computer Science, November 1993,
pp.362-371
\bibitem{[C94]} Claude Cr\'{e}peau. ``Quantum Oblivious Transfer``. Journal of
Modern Optics, 41(12):2455¨C2466, 1994.

\bibitem{[DAUA]} Dorit Aharonov, Amnon Ta-Shma, Umesh V. Vazirani, Andrew Chi-Chih
Yao. ``Quantum bit escrow``. Proceedings of the 32nd Annual ACM
Symposium on Theory of Computing(STOC'00), 2000.

\bibitem{[EGL85]} Even, S., Goldreich, O. and Lempel, A., ``A Randomized Protocol for Signing
Contracts``, Communications of the ACM, vol. 28, pp. 637-647,
1985.
\bibitem{[HO]} W. Hoefding, ``Probability Inequalities for Sums of Bounded Random Variables``, Journal of the American Statistical Association, Vol.58, 1936, pp.13-30
\bibitem{Mayers}    Mayers, D. ``Unconditionally Secure Quantum Bit Commitment is
                Impossible``. Physical Review Letters 78 . pp 3414-3417 (28 April
                1997).
\bibitem{[Naor01]} Moni Naor, Benny Pinkas. ``Efficient Oblivious Transfer Protocols``.  SODA, 2001
\bibitem{[Shor97]}P. W. Shor, ``Polynomial-Time Algorithms for Prime Factorization and Discrete Logarithms on a Quantum Computer``, SIAM Journal on Computing, V.26:(5), 1997.
\bibitem{[R81]} Rabin, M.O., ``How to exchange secrets by Oblivious
Transfer``, technical report TR-81, Aiken Computation Laboratory,
Harvard University, 1981.
\bibitem{[WIE]} Wiesner, S., ``Conjugate coding``, Sigact News,
vol.15, no. 1, 1983,  pp.78-88; Manuscript written circa 1970,
unpublished until it appeared in SIGACT News.
\bibitem{[MJV02]}Yi Mu, Junqi Zhang, Vijay Varadharajan, "m out of n oblivious
transfer," ACISP 2002, Lecture Notes in Computer Science 2384,
Springer Verlag, 2002. pp. 395-405


\end{thebibliography}
\end{document}